# Model Matching Challenge: Benchmarks for Ecore and BPMN Diagrams

Pit Pietsch
Software Engineering Group
University of Siegen
pietsch@informatik.uni-siegen.de

Klaus Müller
Software Engineering
RWTH Aachen University
mueller@se-rwth.de

Bernhard Rumpe
Software Engineering
RWTH Aachen University
rumpe@se-rwth.de

## Abstract

In the last couple of years, Model Driven Engineering (MDE) gained a prominent role in the context of software engineering. In the MDE paradigm, models are considered first level artifacts which are iteratively developed by teams of programmers over a period of time. Because of this, dedicated tools for versioning and management of models are needed. A central functionality within this group of tools is model comparison and differencing.

In two disjunct research projects, we identified a group of general matching problems where state-of-the-art comparison algorithms delivered low quality results. In this article, we will present five edit operations which are the cause for these low quality results. The reasons why the algorithms fail, as well as possible solutions, are also discussed. These examples can be used as benchmarks by model developers to assess the quality and applicability of a model comparison tool for a given model type.

## 1 Motivation

Model Driven Engineering (MDE) has gained a prominent role in the context of software engineering. Within the MDE paradigm, models are first level artifacts and essentially the central development documents. Just like source code, models are typically iteratively developed by teams of programmers over a period of time. Hence, dedicated tools for versioning and management of models are needed [1, 2] to support the developers in their daily routine. Some configuration management tools for models, e.g. Amor [3], were introduced in the last couple of years. A central functionality within this group of tools is model comparison and differencing. A large number of algorithms which implement this function and which are used in different contexts have been proposed recently [4, 5, 6]. In this article, we use the term model differencing only in the sense of syntactic model differencing, which aims at finding structural changes within models. Other definitions to model differencing, e.g. semantic model differencing [7, 8] and respective approaches [9, 10], are not in the scope of this article.

Available model comparison algorithms [11, 12] work reasonably well on class diagrams and similar model types. Support for other model types, e.g. state machines [13] and process models [14], is currently in the focus of research. Still, we were able to identify a group of general matching problems where state-of-the-art model comparison tools deliver differences which are of low quality and sometimes are even unusable from the perspective of the model developer. Generally spoken, the quality of a difference is poor if corresponding elements, which are considered "the same", are not detected, i.e. they are reported by the algorithm as deleted and added, or if inappropriate elements are matched. The quality of a difference delivered by a comparison algorithm is mainly dependent on the computed matching, i.e. the set of correspondences.

Because of this, it is very important for model developers to be aware of the requirements provided by the given modeling domain and the restrictions of available model comparison algorithms.

Therefore, we will discuss five problematic scenarios in which model comparison approaches either delivered low-quality results or where the delivered results lead to dissent amongst the involved model developers. For each case the reasons why the results are of low quality as well as possible solutions are also discussed. All of the examples given in this article originate from two research projects. In one of the projects structural Ecore diagrams where used, while the other project focused on process models. Each of the five problems is presented by example for each model type, therefore it is easy to see that these are general problems which can be transferred to many different model types and domain specific languages.

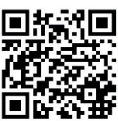

For reasons of confidentiality, we cannot show the original examples. Instead, we restructured and simplified the models as benchmarks in order to focus on the core of the problems.

The set of Ecore examples originate from a joined project of RWTH Aachen University and an industrial partner in which model differencing is performed on UML class diagrams. These UML class diagrams are used in a model-based development project to generate a core part of a complex software system. Due to this, it is crucial to understand how the models changed. The simplification of this task was originally the main motivation to introduce a model differencing algorithm. The differences identified in the model differencing process are analyzed by the modeling experts before the generator is executed to generate the different parts of the software system.

The Business Process Model and Notation Version 2.0 diagrams (BPMN2) which are discussed in this article originate from a research cooperation between the University of Siegen and the Technical University of Dortmund. The question behind this project was whether or not it is possible to identify which security-related constraints on the diagrams had to be re-evaluated based on the difference between two models. The core idea is that usually the difference between two revisions of a model is small, i.e. only few changes are applied. Hence, a large part of the model is not affected and therefore many of the constraints are still valid. Obviously, this approach is very dependent on the quality of the computed differences, i.e. they always have to be correct.

In both projects, commonly used state-of-the-art model comparison algorithms were applied. The differences which these algorithms delivered were sometimes of a low quality and therefore not usable in the context of the projects. We will now discuss five of the edit operations where the algorithms failed to produce high-quality differences or the users could not agree on what can be considered as a high-quality difference in the first place.

The five edit operations[1] are:

- Move Element,

- Rename Element,

- Move Renamed Element,

- Exchange Location of Elements,

- Update Target of Reference/Flow Element.

Each operation is discussed in a section of its own and is presented for Ecore as well as for BPMN2 diagrams. The article ends in Section 7 with a summary and conclusion.

---

[1] All examples can be downloaded at [15].

## 2 Move Element

The first example of a problematic edit operation is *moveElement*. In cases where model elements are moved within the hierarchy of the given model, state-of-the-art differencing algorithms often compute a matching which is generally considered as incorrect by the model developers.

**Move Element for Ecore** The initial version of the Ecore file considered in this first example is shown in Figure 1(a) and it merely contains a single class `DomesticAnimal` with two attributes.

Based on this initial version, we perform two edit operations. First of all, we create a new subpackage `shop` within package `de`. Furthermore, we move the class `DomesticAnimal` into this subpackage. The Ecore file resulting from these edit operations is shown in Figure 1(b).

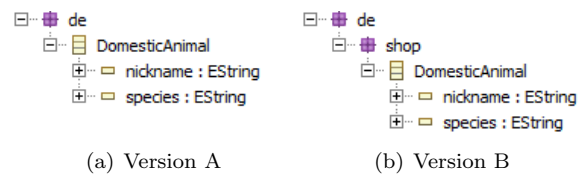

(a) Version A    (b) Version B

Figure 1: Move Model Elements for Ecore

**Move Element for BPMN** At its core version A and version B from Figure 4, both model a trivial process consisting of only one task named `Deliver Goods`. In this example, two edit operations are applied to the original model version A: at first a new subprocess called `Send Order` is inserted and in a second step the existing process is moved into the newly created subprocess.

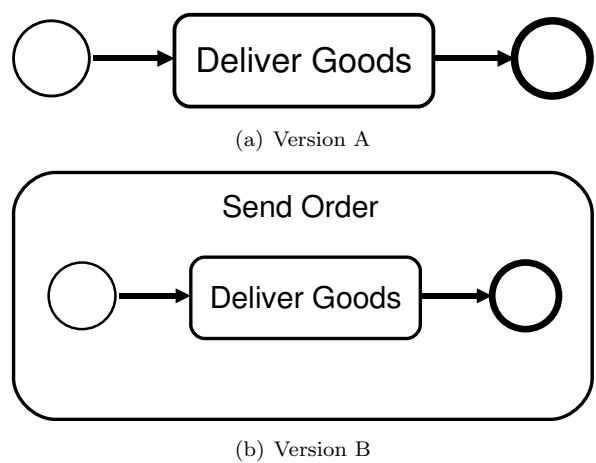

(a) Version A

(b) Version B

Figure 2: Move Model Elements for BPMN

**Expected and Actual Results** Model developers generally expect changes as discussed in Figure 1 and 2 to be reported as move operations. There was some discussion amongst the BPMN developers whether the move of the process should be reported as five atomic operations, i.e. *moveStartEvent*, *moveEndEvent*, *moveTask* and 2x *moveSequenceFlow*, or, as originally intended as one single, complex operation *moveProcessToSubprocess*[2]. In any case, developers still agreed that such changes are best described as moves of the respective model elements.

However, state-of-the-art model comparison approaches often apply a runtime optimization technique called top-down matching. Algorithms that are using top-down matching are not considering all elements of the modified model when they are searching for a possible correspondence for a given element of the original model. Instead they limit the search space by only considering elements as candidates when their parent elements are already matched. While this speeds up the correspondence computation significantly, it is easy to see that these algorithms are not able to detect move operations. Instead they report the elements as deleted from the original and inserted in the modified model. This is obviously in contrast to what model developers would expect.

This unexpected result can be addressed in two different ways. The first approach is to only use algorithms that do not limit the scope of comparison but explicitly search for moved elements in the two models. This has the drawback of longer runtimes when it comes to difference computation. The other approach is to perform a top-down based matching first and apply a matching algorithm which is able to detect moves only on the elements that are not matched after the first phase. The latter approach has the advantage that the search space is already limited by the calculated correspondences when the more complex model comparison algorithm is applied in phase two. In most cases, this offers the best trade-off between performance and quality.

## 3 Rename Element

The second example of an edit operation where model comparison algorithms may report unexpected results is the operation *renameElement*.

**Rename Element for Ecore** For this example, we use the same initial version as in the previous example. This time, we perform two edit operations on the initial version shown in Figure 3(a): at first we rename the class `DomesticAnimal` into `Pet` and in a second step we rename the attribute `nickname` into `moniker`. The resulting Ecore file is shown in Figure 3(b).

**Rename Element for BPMN** Figure 4 shows two versions of a simple process consisting of only a single task. Between version A and version B, there are no structural changes, i.e. no edit steps were applied which add, delete or move elements. Instead only the name of the task `Deliver Goods` from the original Version A has been changed to `Send Items` in the modified Version B.

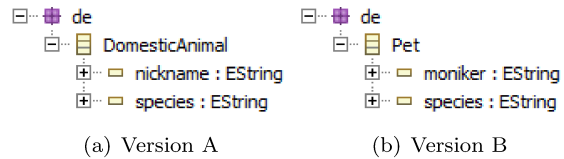

(a) Version A    (b) Version B

Figure 3: Rename Model Elements for Ecore

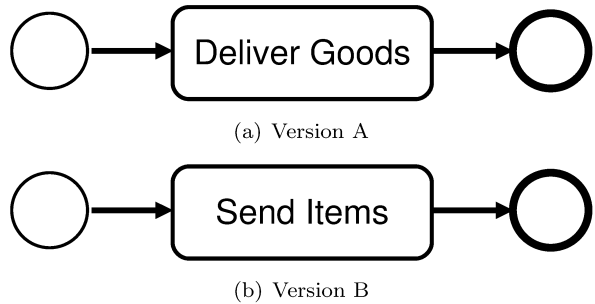

(a) Version A

(b) Version B

Figure 4: Rename Model Element for BPMN

**Expected and Actual Results** For most types of models, names are a very important feature when the correspondences between elements of two models are computed. Ecore as well as BPMN diagrams belong in this category. Almost all elements in such models are named and usually the names are even unique. Most developers expect for the given examples that the respective elements are identified as corresponding and a rename operation is reported by the differencing algorithm. Their reasoning is that they intuitively understand that the meaning of the names in both cases is very similar, and hence they expect this to be reflected by the reported results of a differencing tool. This however is not the case.

While many state-of-the-art model comparison technologies are heavily reliant on the names of model elements, the similarity between two names is usually based only on properties like the length of the string and the contained letters. Some model comparison tools use algorithms like the *Longest Common Subsequence* (LCS) to compute similarities between strings, but such algorithms have drawbacks. The LCS for example delivers low-quality results when a name consists of several words and the sequence of these words is changed. This sometimes occurs in model types like class diagrams, where names often consist of several concatenated parts. Therefore, some algorithms use new, specialized functions when names are compared. One example for this is the algorithm proposed by Xing et al. [11] where the authors introduce a similarity function[3] based on adjacent characters of strings.

---

[2]For algorithms that are able to lift atomic edit operations to a higher abstraction level see [16] or [17].

[3]This function is also used in the EMFCompare project [12], which is based on the work of Xing.

Still, even these improved similarity functions do not recognize the semantics of words and therefore do not report the name of the elements in the given examples in Figure 3 and 4 as updated. Instead they report the deletion of the respective element in the original model and the insertion of a new element in the modified model. Naturally, such a matching is usually considered as incorrect by the model developers.

One way to address this problem is to add new heuristics to model matching algorithms. These functions should not only take the different letters of a string into account but also check linguistic databases, e.g. WordNet [18], to compare the similarity of the semantics of the strings when correspondences are computed. Unfortunately, the authors are not aware of any model comparison algorithm which currently supports such semantic comparison functions for strings.

## 4 Move Renamed Element

A further potentially problematic edit operation is *moveRenamedElement*, i.e. a combination of the edit operations *moveElement* and *renameElement* already presented in the previous sections.

**Move Renamed Element for Ecore** As this scenario is a combination of the previously presented scenarios, we use a combination of the existing examples here. As our initial version, we use the initial version from the previous two examples again. In addition to moving the class `DomisticAnimal` into the new package `shop`, we also rename the moved class `DomisticAnimal` into `Pet` and the attribute `nickname` into `moniker`. The Ecore file resulting from these edit operations is shown in Figure 5(b).

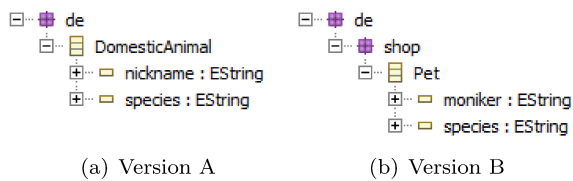

(a) Version A             (b) Version B

Figure 5: Move Renamed Model Element for Ecore

**Move Renamed Element for BPMN** For this example, the two edit operations already discussed in Section 2 and Section 3 are applied to version A of the BPMN diagram. At first the task `Deliver Goods` is renamed to `Send Items`. Then the subprocess `Send Order` is created and the original process is moved into the new subprocess.

**Expected and Actual Results** Usually, an end-user would expect the move and the renaming of the elements to be reported. However, as already pointed out in the previous sections, some state-of-the-art model comparison approaches search for matching elements in a limited scope only. Furthermore, many approaches heavily rely on the name of elements. These two observations combined can induce that the move and rename operations are not detected properly but delete and insertion operations are reported instead.

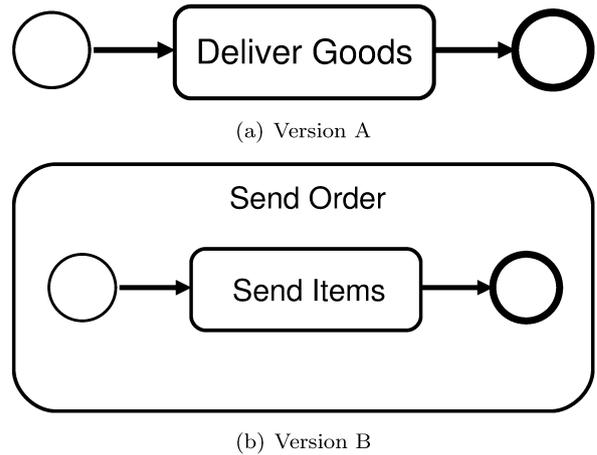

(a) Version A

(b) Version B

Figure 6: Move Renamed Model Element for BPMN

This problem can be addressed by applying a combination of the previously introduced proposals. This means that a model matching algorithm should not only take the semantics of the names into account but also either search for matching elements in the whole models or perform a top-down based matching first and apply a matching algorithm on the unmatched elements after that.

## 5 Exchange location of elements

The next edit operation we will discuss is *exchangeElementLocation*. This edit operation comprises two *moveElement* operations, as one element is moved to the location of the other and vice versa.

**Exchange Location for Ecore** In this example, we use the Ecore file shown in Figure 7(a) as our initial version. It only contains two different classes `DomisticAnimal` and `DomisticAnimalNew` with the same attributes that are located in different packages.

Starting from this initial version, we perform two edit operations. On the one hand, we move the class `DomisticAnimal` into the package `shop`. Furthermore, we move the class `DomisticAnimalNew` into the package `core`. In that way, we exchange the location of both classes. The Ecore file resulting from these edit operations is shown in Figure 7(b).

**Exchange Location for BPMN** The according example of the *exchangeLocation* edit operation for BPMN diagrams can be seen in Figure 8. Here, two move operations are applied. The first operation moved the task `doSomething` from subprocess `Left` to subprocess `Right` while the second edit operation moved the task `doSomethingNew` from subprocess `Right` to `Left`. This kind of edit operation, where nearly clones of elements are moved, are rather uncommon and are not expected to occur in real models

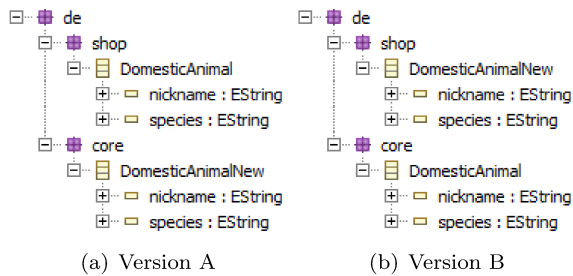

(a) Version A   (b) Version B

Figure 7: Exchange Elements for Ecore

on a regular basis.

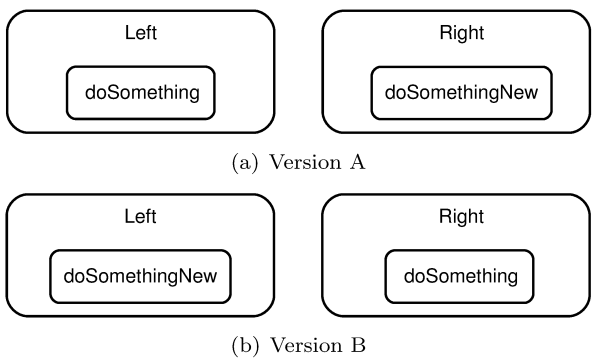

(a) Version A

(b) Version B

Figure 8: Exchange Elements for BPMN

**Expected and Actual Results** Discussions with users revealed that the opinions concerning the correct matching differ heavily. Some users argued that this scenario should be regarded as having performed rename operations. On the other hand, other users expected move operations as a result. This scenario illustrates very well that there can be different interpretations of what kinds of changes were performed, even if the changes were very small.

There are basically two aspects that influence which change operations are reported by a model comparison technique in this scenario. As already stated in Section 2, some model comparison techniques search for matching elements in a limited scope only. Consequently, these approaches search for the best matching element in a limited scope only. Thus, it can happen that a matching element is found, albeit in another scope a further, potentially better, matching element exists which is then not considered. As a result, a potentially suboptimal matching is computed, whereas in the other scenario potentially no matching is computed. In addition, a model comparison approach might attach difference importance to specific edit operations. It might, e.g., rather report a move operation than a rename operation.

One possible approach to address the first problem is to search for (the best) matching elements in the whole model, as already outlined in Section 2. The downside of applying this strategy is the deterioration of the runtime of the difference computation. One approach to cope with the second problem is to let the users influence which edit operations have what importance or in which situations a change of a certain type should be reported.

## 6 Update Reference Target

The last edit operation which is discussed in this article is *updateReferenceTarget*. The problem with this edit operation is not rooted in the model comparison algorithms themselves but comes from the way developers interpret the operation. While some developers expect the change to be reported as the actual update of the reference target, other developers expect it to be reported as a deletion and insertion of the reference itself, arguing that this change severely changes the semantic of the element.

**Update Reference Target for Ecore** The initial version for our last example is given in Figure 9(a). It contains three classes and one association from `DomisticAnimal` to `Owner`.

This version is now changed in such a way that the target of the existing association does no longer reference `Owner`, but `Person` instead, with the result shown in Figure 9(b).

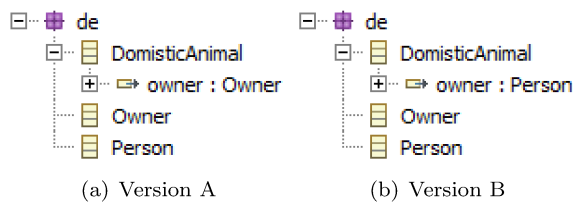

(a) Version A   (b) Version B

Figure 9: Update Reference Target

**Update Sequence Flow Target for BPMN** The equivalent for *UpdateReferenceTarget* in the context of BPMN diagrams is depicted in Figure 10, where the target of the sequence flow has been changed from *Task2* to *Task3*.

**Expected and Actual Results** While the actual edit operation was identified by most model comparison algorithms, there were users which considered this result as incorrect. Instead they argued that the change of the target of a sequence flow fundamentally changed the meaning of this element for the model and should rather be treated as an insertion and deletion of the sequence flow.

Obviously, there is no clear-cut solution which can address both perspectives. One approach of addressing this problem is to implement different algorithms for the different expectations of the users, which is very laborious. Another approach is to use a configurable model comparison algorithm and to make the detection of the target change optional. This means that users who like to think of the edit operation as a simple update of the target reference get this result re-ported, while other users get any change of the source or target of a sequence flow reported as a deletion and insertion of a sequence flow. For this, a model comparison algorithm has to be configurable in a way that it either reports references as corresponding only if neither the source nor the target has been changed, or, alternatively, references can correspond when they have the same target. An example of such an adaptable model comparison algorithm is discussed in [13]. Interestingly, there was a general consensus between model developers that an update of the source of a reference should always be reported as a delete and an insert operation. This is another example that personal preferences as well as the semantics which are given to edit operations by the developers are of major importance when differences are computed.

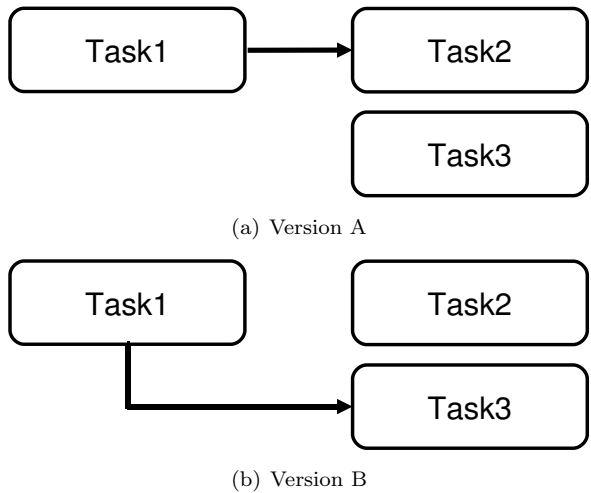

Figure 10: Update Sequence Flow Target

## 7 Conclusion

In this article, we presented multiple examples for change scenarios of models, which can pose problems for state-of-the-art model comparison techniques. These examples can be used to identify restrictions and implicit assumptions of model comparison techniques, as these are often not clear to a potential user of a model comparison tool. Furthermore, these examples can be used to assess to what extent the user can adapt the model comparison technique to own requirements.

Off-the-shelf products usually work reasonably well for typical use cases and for particular model types such as class diagrams or similar model types. Deficiencies of these tools could be overcome by using special-purpose comparison tools which focus on comparing selected model types. However, the development and maintenance of such tools is expensive, especially for domain specific languages that are not widespread or for meta models that change rapidly. Hence, we need model comparison tools which can be adapted to a specific model type, user preferences or the application context [13] in order to be able to cope with these insufficiencies appropriately.

For future work, it is planned to design a complete *Model Matching Challenge* (MMC) which will contain the benchmarks given in this article as well as additional scenarios. Such a MMC can be used to assess the quality of model comparison algorithms. It is intended to design the MMC in order to demonstrate strength and weaknesses of state-of-the-art model comparison algorithms so that end-users can better understand which algorithms are suited for a specific model type or application context.

## References

[1] T. Mens, M. Wermelinger, S. Ducasse, S. Demeyer, R. Hirschfeld and M. Jazayeri, "Challenges in software evolution," in *Proc. International Workshop on Principles of Software Evolution (IWPSE'05)*, 2005, pp. 13–22.

[2] T. Levendovszky, B. Rumpe, B. Schätz, and J. Sprinkle, "Model evolution and management," in *Proc. International Dagstuhl conference on Model-based engineering of embedded real-time systems (MBEERTS'07)*, 2007, pp. 241–270.

[3] Amor Adaptable Model Versioning homepage; http://www.modelversioning.org, [Accessed 11-May-2013].

[4] Bibliography on Comparison and Versioning of Software Models; http://pi.informatik.uni-siegen.de/CVSM/, [Accessed 11-May-2013].

[5] D. Kolovos, D. Di Ruscio, A. Pierantonio, and R. Paige, "Different models for model matching: An analysis of approaches to support model differencing," in *Proc. International Workshop on Comparison and Versioning of Software Models (CVSM'09)*, 2009, pp. 1–6.

[6] M. Stephan and J. R. Cordy, "A Survey of Methods and Applications of Model Comparison," in *Technical Report 2011-582 Rev. 3, School of Computing, Queen's University*, June 2012.

[7] S. Maoz, J.O. Ringert and B. Rumpe, "A Manifesto for Semantic Model Differencing," in *Proc. International Workshop on Models and Evolution (ME'10)*, 2010, pp. 194–203.

[8] S. Maoz, J.O. Ringert and B. Rumpe, "An Interim Summary on Semantic Model Differencing," in *Softwaretechnik-Trends, Volume 32, Issue 4*, 2012.

[9] S. Maoz, J.O. Ringert and B. Rumpe, "CDDiff: Semantic Differencing for Class Diagrams," in *Proc. European Conference on Object Oriented Programming (ECOOP'11)*, 2011, pp. 230–254.

[10] S. Maoz, J.O. Ringert and B. Rumpe, "ADDiff: Semantic Differencing for Activity Diagrams," in *Proc. of the 19th ACM SIGSOFT symposium and the 13th European conference on Foundations of software engineering (ESEC/FSE'11)*, 2011, pp. 179–189.

[11] Z. Xing and E. Stroulia, "UMLdiff: An algorithm for object-oriented design differencing," in *Proc. International Conference on Automated software engineering (ASE'05)*. 2005, pp. 54–65.

[12] "EMF Compare Project," `http://www.eclipse.org/emf/compare`, 2012, [Accessed 11-May-2013].

[13] T. Kehrer, U. Kelter, P. Pietsch, and M. Schmidt, "Adaptability of Model Comparison Algorithms," in *Proc. International Conference on Automated software engineering (ASE'12)*, 2012, pp. 306–309.

[14] P. Pietsch, S. Wenzel, "Comparison of BPMN2 Diagrams," in *International Workshop on the Business Process Model and Notation (BPMN'12)*, 2012, pp. 83–97.

[15] `http://se-rwth.de/materials/mmc/EcoreBPMN2.zip`, [Accessed 11-May-2013].

[16] P. Langer, "Adaptable Model Versioning based on Model Transformation By Demonstration", PHD Thesis, TU Wien, 2011.

[17] T. Kehrer, U. Kelter, G. Taentzer, "A rule-based approach to the semantic lifting of model differences in the context of model versioning," in *Proc. International Conference on Automated software engineering (ASE'11)*, 2011, pp. 163–172.

[18] "WordNet Lexical Database," http://wordnet.princeton.edu/, 2012, [Accessed 11-May-2013].